%% file: main.tex
\documentclass[10pt,conference]{IEEEtran}
\IEEEoverridecommandlockouts

\PassOptionsToPackage{table,xcdraw}{xcolor}
\usepackage{colortbl}
\usepackage{tcolorbox}
\usepackage{tikz}
\usepackage{mathdots}
\usetikzlibrary{fadings}
\usetikzlibrary{patterns}
\usetikzlibrary{shadows.blur}
\usetikzlibrary{shapes}
\usetikzlibrary{shapes.geometric, arrows.meta, positioning}
\usepackage{listings}
\usepackage{array}
\usepackage{makecell}
\usepackage{siunitx}
\usepackage{multirow}
\usepackage{tabularx}
\usepackage{extarrows}
\usepackage{booktabs}
\usepackage{enumitem}
\usepackage{titlesec}
\usepackage{caption}
\usepackage{textcomp}
\usepackage{url}
\usepackage{hyperref}

\usepackage{cite}
\usepackage{amsmath,amssymb,amsfonts}
\usepackage{algorithmic}
\usepackage{graphicx}
\usepackage{textcomp}
\usepackage{xcolor}
\def\BibTeX{{\rm B\kern-.05em{\sc i\kern-.025em b}\kern-.08em
    T\kern-.1667em\lower.7ex\hbox{E}\kern-.125emX}}

\newcommand{\tool}[0]{\mbox{\textsc{InitScope}}}
\newcommand{\bench}[0]{\mbox{\textsc{SCABench}}}

\newcommand{\lstbg}[3][0pt]{{\fboxsep#1\colorbox{#2}{\strut #3}}}
\lstdefinelanguage{diff}{
  basicstyle=\footnotesize \ttfamily \color{black},
  columns=fullflexible,
  breaklines=true,
  breakatwhitespace=false,
  showspaces=false,               
  showstringspaces=false,  
  frame=single, 
  showtabs=false,
  numbersep=5pt,
  showstringspaces=false,        
  stepnumber=1,                   
  tabsize=5,                     
  title=\lstname,  
  numbers=left,                 
  numbersep=5pt,   
  backgroundcolor=\color{white},
  morecomment=[f][\lstbg{red!30}]-,
  morecomment=[f][\lstbg{green!30}]+,
  morecomment=[f][\textit]{@@},
  moredelim=**[is][\color{red}]{@}{@},
  moredelim=**[is][\color{green}]{##}{##},
}
\definecolor{white}{rgb}{0.98,0.98,0.98}
\definecolor{dkgreen}{rgb}{0,0.6,0}
\definecolor{dred}{rgb}{0.545,0,0}
\definecolor{dblue}{rgb}{0,0,0.545}
\definecolor{lgrey}{rgb}{255,0.9,0.9}
\definecolor{gray}{rgb}{0.4,0.4,0.4}
\definecolor{darkblue}{rgb}{0.0,0.0,0.6}

\definecolor{pblue}{rgb}{0.13,0.13,1}
\definecolor{pgreen}{rgb}{0,0.5,0}
\definecolor{pred}{rgb}{0.9,0,0}
\definecolor{pgrey}{rgb}{0.46,0.45,0.48}
\lstdefinestyle{prStyle}{
  showspaces=false,
  showtabs=false,
  breaklines=true,
  showstringspaces=false,
  breakatwhitespace=true,
  postbreak=\mbox{\textcolor{black}{$\hookrightarrow$}\space},
  captionpos=b,    
  commentstyle=\color{pgreen},
  keywordstyle=\color{pblue},
  stringstyle=\color{pred},
  basicstyle=\ttfamily,
  moredelim=[il][\textcolor{pgrey}]{},
  frame=tb,
  moredelim=[is][\textcolor{pgrey}]{\%\%}{\%\%}
  moredelim=**[is][\color{red}]{@}{@},
  moredelim=**[is][\color{dkgreen}]{##}{##},
}
\lstdefinelanguage{cpp}{
  backgroundcolor=\color{white},  
  basicstyle=\scriptsize \ttfamily \color{black} ,   
  breakatwhitespace=false,       
  breaklines=true,
  postbreak=\mbox{\textcolor{black}{$\hookrightarrow$}\space},
  captionpos=b,                   
  commentstyle=\color{dkgreen},   
  deletekeywords={...},          
  escapeinside={\%*}{*)},                  
  frame=single,                  
  language=C++,                
  keywordstyle=\color{dblue},  
  morekeywords={BRIEFDescriptorConfig,string,TiXmlNode,DetectorDescriptorConfigContainer,istringstream,cerr,exit}, 
  identifierstyle=\color{black},
  stringstyle=\color{blue},      
  numbers=right,                 
  numbersep=5pt,                  
  numberstyle=\tiny\color{black}, 
  rulecolor=\color{black},        
  showspaces=false,               
  showstringspaces=false,        
  showtabs=false,                
  stepnumber=1,                   
  tabsize=5,                     
  title=\lstname,
  moredelim=**[is][\color{red}]{@}{@},
  moredelim=**[is][\color{dkgreen}]{##}{##},
}

\usepackage{amsthm}

\tcbset{
  Customquote/.style={               
    colback=gray!10,                  
    colframe=black!70,                
    boxrule=0.5mm,                    
    arc=4mm,                          
    auto outer arc,                   
    boxsep=5pt,                       
    left=5pt, right=5pt,              
    top=5pt, bottom=5pt               
  }
}

\newtcolorbox{customquote}{Customquote}

\newtcolorbox{participantquote}{
  colback=gray!10,
  colframe=black!40,
  boxrule=0.3pt,
  arc=2pt,
  left=6pt,
  right=6pt,
  top=4pt,
  bottom=4pt,
  fontupper=\itshape,
  before skip=6pt,
  after skip=6pt
}

\usepackage{wasysym}
\newcommand{\fullmark}{\checkmark}
\newcommand{\nonemark}{$\times$}
\newcommand{\partialmark}{\raisebox{-0.15ex}{\scalebox{0.9}{\LEFTcircle}}}

\begin{document}

\title{Cold-Start Anti-Patterns and Refactorings in Serverless Systems: An Empirical Study
\thanks{This work has been submitted to the IEEE for possible publication. Copyright may be transferred without notice, after which this version may no longer be accessible. Accepted at IEEE SANER 2026. This is the author preprint; camera-ready version forthcoming.}
}


\author{
\IEEEauthorblockN{
Syed Salauddin Mohammad Tariq\IEEEauthorrefmark{1},
Foyzul Hassan\IEEEauthorrefmark{1},
Amiangshu Bosu\IEEEauthorrefmark{2},
Probir Roy\IEEEauthorrefmark{1}
}
\IEEEauthorblockA{\IEEEauthorrefmark{1}University of Michigan--Dearborn,
Dearborn, Michigan, USA\\
\{ssmtariq, foyzul, probirr\}@umich.edu}
\IEEEauthorblockA{\IEEEauthorrefmark{2}Wayne State University,
Detroit, Michigan, USA\\
amiangshu.bosu@wayne.edu}
}

\maketitle

\begin{abstract}
Serverless computing simplifies deployment and scaling, yet cold-start latency remains a major performance bottleneck. 
Unlike prior work that treats mitigation as a black-box optimization, we study cold starts as a \emph{developer-visible design problem}. 
From 81 adjudicated issue reports across open-source serverless systems, we derive taxonomies of initialization anti-patterns, remediation strategies, and diagnostic challenges spanning design, packaging, and runtime layers. 
Building on these insights, we introduce \bench{}, a reproducible benchmark, and \tool{}, a lightweight analysis framework linking what code is \emph{loaded} with what is \emph{executed}. 
On \bench{}, \tool{} improved localization accuracy by up to $40\%$ and reduced diagnostic effort by $64\%$ compared with prior tools, while a developer study showed higher task accuracy and faster diagnosis. 
Together, these results advance evidence-driven, performance-aware practices for cold-start mitigation in serverless design.

\noindent\textbf{Availability:} The research artifact\footnote{Artifact available at \url{https://doi.org/10.5281/zenodo.17375894}} 
is publicly accessible for future studies and improvements.

\end{abstract}

\begin{IEEEkeywords}
Serverless computing, Performance anti-patterns, Cold-Start, Software development tools.
\end{IEEEkeywords}

\input{sections/1-introduction}
\input{sections/2-background}

\input{sections/3-approach_v2}
\input{sections/4-Evaluation}

\input{sections/7-threats-to-validity}

\input{sections/8-Conclusions}
\input{sections/9-references}

\end{document}

%% file: sections/1-introduction.tex
\section{Introduction}
\label{sec:intro}

Serverless computing has emerged as a widely adopted paradigm that simplifies software deployment by abstracting infrastructure management while providing automatic scaling and fine-grained billing~\cite{li2022serverless,shafiei2022serverless,adzic2017serverless}. Despite these advantages, serverless applications frequently experience unpredictable latency caused by \textit{cold starts}, which occur when an inactive function must reinitialize its execution environment before handling a new request~\cite{vahidinia2020cold,solaiman2020wlec}. Even short delays during cold starts can significantly affect user experience and business outcomes. For instance, Amazon reported a 1\% loss in sales for every additional 100 ms of latency, and Google observed a 20\% drop in search traffic when response time increased by 500 ms~\cite{amazon_study,google_study}. 

Cold-start latency originates from multiple sources that span both cloud platforms and application code. While platform-level provisioning delays, such as container scheduling and runtime sandbox initialization, are largely managed by cloud providers, application-level inefficiencies remain under developer control. These inefficiencies, which cause performance overhead rather than correctness defects, include redundant library imports, deferred one-time initialization, and unnecessary packaging of large dependencies. Despite their practical impact, the software engineering community lacks a systematic understanding of how such inefficiencies arise, how often they occur in practice, and how developers can effectively identify and address them.

Prior work has primarily focused on \emph{automated mitigation}. Techniques such as \textit{FaaSLight}~\cite{liu2023faaslight} remove unused dependencies through static analysis, while \textit{SLIMStart}~\cite{slimstart} applies adaptive profiling and lazy loading to reduce startup latency.
Although these approaches achieve measurable gains, they often treat inefficiencies as black-box optimization targets and provide limited insight into the \emph{developer-level causes and reasoning} behind them. Moreover, they may generate suboptimal and inaccurate suggestions when applied to complex code. Such inaccuracy, combined with their black-box nature, hinders the adoption of these approaches in real-world scenarios.
In practice, cold-start problems extend across multiple abstraction layers, from import graphs and packaging to runtime and environment, which makes them difficult to isolate and address in a unified manner.
Addressing these challenges requires collaboration within the software engineering community: researchers can model the underlying mechanisms, tool builders can translate them into diagnostic methods, and practitioners can apply and refine these optimizations in real-world systems. Accordingly, this lack of developer-centered understanding motivates an empirical investigation into the root causes, remediation strategies, and diagnostic challenges surrounding initialization inefficiencies in serverless systems.

To address this knowledge gap, we develop a unified empirical model of \textbf{developer-visible initialization inefficiencies} in serverless and cloud-native systems.
The model integrates three analytical dimensions:
(1)~\emph{Anti-patterns:} recurring code-, configuration-, and packaging-level mechanisms that introduce cold-path overheads;
(2)~\emph{Optimization strategies:} developer and tooling practices that mitigate or restructure these inefficiencies; and
(3)~\emph{Localization challenges:} factors that impede the attribution, reasoning, and diagnosis of initialization costs across layers.
Together, these dimensions form a \emph{cause-mitigation-diagnosis continuum} linking developer actions, tool support, and performance outcomes. The research questions guiding this study are introduced in Section~\ref{sec:rqs}.

For the study, we mined 226 issue reports from 35 open-source serverless and cloud-native repositories and adjudicated 81 of them in depth.  
Using grounded coding~\cite{vollstedt2019introduction}, we derive a \textbf{root-cause taxonomy of initialization anti-patterns}, a complementary \textbf{refactoring taxonomy of optimization strategies}, and a set of \textbf{diagnostic challenges} that capture localization difficulty.  
We then instantiate these patterns in \bench{}, a benchmark suite of reproducible cold-start scenarios, and assess detection support using the proposed approach, \tool{} and other existing tools.

Empirically, we find that cold-start inefficiencies are \textbf{systemic rather than incidental}: in our dataset of 81 adjudicated issues, over 80\% stem from recurring cross-layer mechanisms that expand dependency scope or defer initialization to the request path. Remediation practices converge on a reproducible \textbf{refactoring lifecycle}, that is reducing dependency scope, relocating initialization to shared or early phases, and hardening packaging and runtime settings for stability. These patterns recast cold-start mitigation as an \textbf{architectural design concern}, emphasizing explicit cold-path boundaries, minimal dependency surfaces, and reusable initialization components. Yet, diagnosis remains limited: missing import-phase signals and cold-path visibility gaps frequently obscure causality, underscoring the need for analysis tools that reveal the mapping between what is loaded and what is executed.

In summary, this paper makes the following contributions:

\begin{itemize}[leftmargin=*]
  \item A \textbf{unified empirical model} connecting inefficiency mechanisms, optimization strategies, and localization challenges across design, packaging, runtime, and environment layers.
  \item Two complementary \textbf{taxonomies}, a root-cause taxonomy of initialization anti-patterns and a remediation taxonomy of optimization strategies, derived from 81 adjudicated issues.
  \item \textbf{\bench{},} a benchmark suite and dataset supporting reproducible evaluation of cold-start inefficiencies and diagnostic tooling.
  \item \textbf{\tool{},} a lightweight analysis framework capturing import-execution mappings, enabling empirical characterization and cross-tool comparison.
  \item A comparative \textbf{developer study} showing that usage-aware diagnostics increased localization accuracy from $14\%$ to $71\%$ and reduced diagnostic time by up to $64\%$.
\end{itemize}

All datasets, scripts, and benchmarks will be released as part of the \textit{\bench{}} artifact package to support independent replication.

The remainder of this paper is organized as follows. 
Section~\ref{background} reviews background and related work. 
Section~\ref{methodology} introduces the research questions and the four-phase empirical framework, detailing data collection, issue mining, taxonomy construction, benchmarking, and evaluation design. 
Section~\ref{eval} reports the empirical findings for each research question. 
Section~\ref{discussion} discusses implications and broader reflections. 
Section~\ref{threats} outlines the main threats to validity and mitigation strategies. 
Finally, Section~\ref{conclusions} concludes the paper and highlights directions for future research.

%% file: sections/2-background.tex
\section{Background and Related Work}
\label{background}

\subsection{Cold-Start Latency: Application vs. Platform Causes}

Cold starts in serverless systems occur when idle functions are reinitialized, incurring delays from both infrastructure and application layers. While platform optimizations such as snapshotting, microVMs, and image caching reduce provisioning time~\cite{vahidinia2020cold,solaiman2020wlec}, recent studies show that \emph{application-layer overheads increasingly dominate}~\cite{liu2023faaslight,warmswap}. In Python-based workloads, importing large libraries such as NumPy or PyTorch can account for over 90\% of cold-start latency~\cite{warmswap}, exceeding container startup costs. These delays arise from top-level imports, runtime configuration, or static setup code, which are regions fully controlled by developers and are often opaque to current toolchains.

\subsection{Developer Practices and Diagnostic Gaps}

Cold-start inefficiencies often emerge from cross-layer interactions spanning design, packaging, and runtime logic, extending beyond platform-level causes. While optimization tools such as FaaSLight~\cite{liu2023faaslight} and SlimStart~\cite{slimstart} reduce startup latency through static or profile-guided trimming, they treat inefficiencies as black-box performance problems and provide little insight into how such issues arise in real-world code. Similarly, general-purpose profilers (e.g., \texttt{cProfile}~\cite{cprofile}, py-spy~\cite{py-spy}, Scalene~\cite{288540}) capture steady-state costs but overlook import-time and initialization behavior.

Our study complements this line of work by shifting focus from automation to understanding: rather than proposing another optimization technique, we empirically characterize \emph{how} initialization inefficiencies emerge, are diagnosed, and are mitigated in practice, providing an explanatory foundation for future analysis tools.

\subsection{Empirical Studies and Taxonomy Gaps}

Empirical work has examined cold-start behavior via platform traces~\cite{shahrad2020serverless}, benchmarks~\cite{kim2019functionbench}, and developer discussions~\cite{wen2021}. Wen et al.~\cite{wen2021} mined Stack Overflow to surface common serverless challenges, including cold starts, but their taxonomy is broad and conceptual, offering little insight into structural causes or remediation behavior. Golec et al.~\cite{golec2024} surveyed cold-start mitigation techniques across system layers but did not explore how inefficiencies emerge or how developers reason about them.

Our work addresses these gaps by constructing a fine-grained taxonomy of cold-start anti-patterns and refactorings from real-world issue reports, and by evaluating profiling tools for their diagnostic support.

%% file: sections/3-approach_v2.tex
\section{Research Questions and Study Design}
\label{methodology}

This section first describes the research questions, followed by the study design, outlining the empirical workflow, data acquisition and analysis procedures, and the methodological practices adopted.

\subsection{Research Questions}
\label{sec:rqs}

We empirically examine how initialization inefficiencies arise, are mitigated, and diagnosed in serverless systems through four guiding research questions.

\textbf{RQ1. What types of initialization \emph{anti-patterns} occur in serverless systems, and how can they be systematically classified across software layers?}  
RQ1 characterizes the underlying causes of initialization inefficiencies beyond platform-level startup behavior, focusing on how design, packaging, runtime, and environment choices in application code contribute to latency.

\textbf{RQ2. What \emph{optimization strategies} mitigate these inefficiencies, and how do they relate to the identified anti-patterns?}  
RQ2 examines remediation practices observed in developer discussions, mapping inefficiency mechanisms (from RQ1) to concrete refactoring and optimization strategies that inform benchmark design.

\textbf{RQ3. What \emph{localization challenges} hinder developers in diagnosing initialization inefficiencies?}  
RQ3 explores why attribution remains difficult, such as hidden import-time work, opaque dependencies, or environment-specific effects, and motivates tool support for improved diagnosis.

\textbf{RQ4. How effectively do existing tools detect and explain these inefficiencies under controlled conditions?}  
RQ4 evaluates profiling and static-analysis tools on the \bench{} benchmark and assesses whether our proposed instrumentation (\tool{}) improves localization accuracy and developer productivity.

\subsection{Empirical Framework}
\label{sec:concepts}

\begin{figure}[t]
  \centering
  \includegraphics[width=\columnwidth]{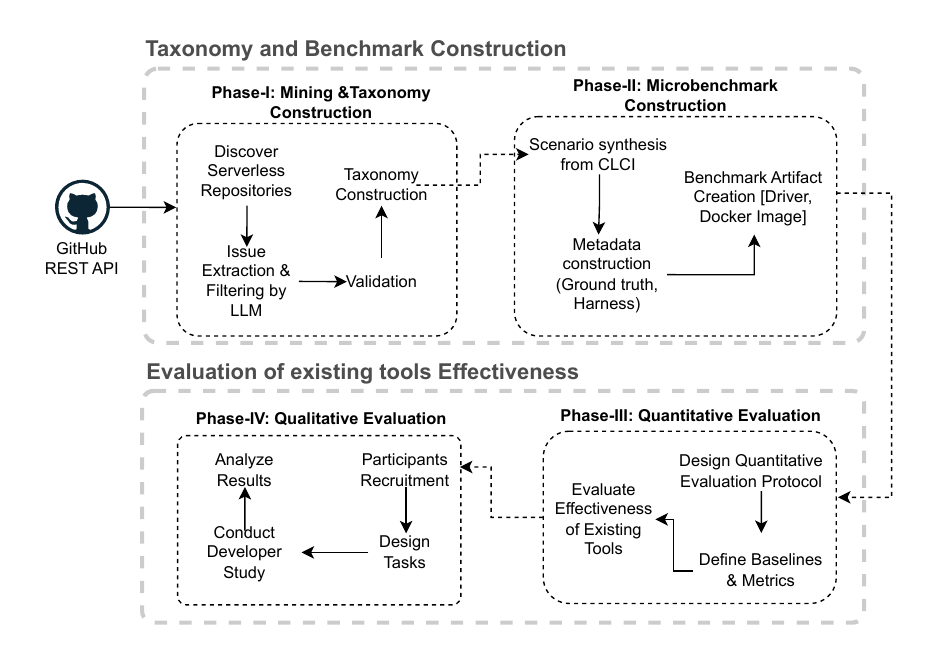}
  \caption{Empirical workflow and RQ mapping.
Phase I: issue mining $\rightarrow$ taxonomies.
Phase II: \bench{} construction.
Phase III: profiling and comparative evaluation on \bench{}.
Phase IV: developer study.}
  \label{fig:workflow}
\end{figure}

Figure~\ref{fig:workflow} presents the four-phase empirical framework integrating qualitative, quantitative, and behavioral analyses to link the \textit{causes}, \textit{mitigation strategies}, and \textit{diagnostic challenges} of cold-start inefficiencies across design, packaging, runtime, and environment layers. Phase~I derives grounded taxonomies of inefficiency mechanisms, optimization strategies, and recurrent \textit{localization challenges} from developer-reported issues; Phase~II instantiates these insights as reproducible benchmark scenarios (\bench{}); Phase~III quantitatively evaluates cold-start localization accuracy with \tool{}, our proposed lightweight analysis framework, in comparison to existing baselines; and Phase~IV complements tool-level findings with a developer study examining diagnostic reasoning. Together, these phases establish a unified empirical foundation connecting inefficiency causes (RQ1), optimization strategies (RQ2), localization challenges (RQ3), and evaluation of analysis tools (RQ4).

\subsection{Phase I - Mining-Driven Taxonomy Construction}
\label{sec:phase1}

We began by identifying serverless repositories and mining cold-start-related issues to construct three complementary perspectives: taxonomies of initialization inefficiency mechanisms and optimization strategies, and a set of \textit{localization challenges} describing developer diagnostic difficulties.

\textbf{Repository Discovery and Labeling.}
We collected candidate repositories via the GitHub REST API using domain-representative keywords (\texttt{serverless}, \texttt{aws-lambda}, \texttt{cloudflare-workers}, \texttt{azure-functions}, \texttt{google-cloud-functions}, \texttt{openfaas}, and \texttt{knative}) appearing in topics, names, descriptions, or README files. These keywords follow established terminology in prior empirical studies on serverless computing~\cite{wen2021,golec2024}. Repositories were required to (i) be public, (ii) show activity within six months, and (iii) have at least $1{,}000$~stars, yielding $1{,}315$ projects spanning frameworks, applications, and infrastructure tools.  
Using the \texttt{gpt-oss:20B} open-weight model~\cite{agarwal2025gpt}, we categorized each repository into one of five roles: (i) Meta (e.g., tutorials or design primers), (ii) Library/Framework, (iii) Hosting Platform, (iv) Tooling/Infrastructure Utility, and (v) Application, then excluded meta repositories. The final dataset comprised 72 repositories used for issue mining, summarized in Table~\ref{tab:repo-summary}.

\begin{table}[t]
\centering
\caption{Issue Mining Pipeline.}
\label{tab:repo-summary}
\begin{tabular}{p{0.70\linewidth}p{0.2\linewidth}}
\toprule
\textbf{Filter / Operation (with Purpose)} & \textbf{Repos Left} \\
\midrule
1.\;Initial retrieval using domain keywords (\texttt{serverless}, \texttt{aws-lambda}, etc.) to collect relevant projects. & 1{,}315 \\
2.\;Active projects (\,$\geq$\,1 commit / 6 mo.) ensuring maintained repositories. & 988 \\
3.\;Popular projects (\,$\geq$\,1{,}000 stars) to retain mature, community-visible repositories. & 212 \\
4.\; Remove meta/tutorial repositories to target technical codebases. & 72 \\
5.\;Identify cold-start issues (2019--2024) via keyword search. & 10{,}589 issues \\
6.\;LLM triage to isolate cold-start-relevant reports. & 226 issues \\
7.\;Manual validation to confirm application-level initialization inefficiencies. & 81 issues / 35 repos \\
\bottomrule
\end{tabular}
\end{table}

\textbf{Issue Extraction and Filtering.}
We searched the issue trackers of these 72 repositories for reports created within the last five years (2019-2024) using a set of cold-start and initialization-related keywords applied to issue titles and descriptions:  
\texttt{cold start}, \texttt{initialization latency}, \texttt{slow start}, \texttt{import time}, \texttt{startup overhead}, and \texttt{init delay}.  
This search yielded 10\,589~issues.
To efficiently identify cold-start–related reports, we employed the \texttt{gpt-oss:20B} model for coarse classification, following recent research that leveraged large language models for issue and defect categorization in software repositories~\cite{colavito2024leveraging, koyuncu2025exploring}. The model was prompted as follows: \textit{“Determine whether this issue describes a cold-start performance problem in a serverless system. Respond with one label: ‘Cold-start related’ or ‘Not related.’”}. This filtering reduced the corpus to 226 issues labeled as cold-start related.


\textbf{Manual Validation and Reliability of LLM-Assisted Filtering.}
Two authors with over five years of experience in software performance analysis manually examined all 226 issues labeled by the LLM-assisted triage to confirm relevance, ensuring that reports described application-level cold-start behaviors rather than provider-level or infrastructure delays.  
This cross-verification demonstrated strong alignment between model predictions and human judgment, with substantial inter-rater agreement (Cohen's~$\kappa{=}0.84$).  
During validation, we retained only those reports that not only mentioned cold-start phenomena but also provided explicit discussion of initialization latency or performance bottlenecks.  
Many of the remaining issues described general deployment delays, configuration mishaps, or environment setup tasks without measurable performance implications, and were therefore excluded.  
The finalized corpus comprises 81 issues across 35 repositories that explicitly reference initialization overhead, dependency loading, or deferred-import inefficiencies, providing a validated foundation for subsequent analysis.

\textbf{Taxonomy Construction.}
Each \emph{issue report} was analyzed as the unit of observation, with metadata collected on repository, components, reproduction steps, proposed mitigation, and discussion activity. 
Following established descriptive coding methods~\cite{saldana2015coding,seaman1999qualitative}, two authors independently coded a pilot subset (40\% of the corpus) to inductively identify cross-layer inefficiencies, optimization strategies, and localization challenges. 
Codes were iteratively refined through constant comparison and analytic memos, yielding a shared codebook with clear definitions, inclusion criteria, and examples. 
Using this codebook, the authors independently coded the remaining issues and consolidated findings into three complementary perspectives: 
(i)~\textit{Cross-Layer Cold-start Inefficiencies}, capturing recurring inefficiency types across design, packaging, runtime, and environment layers; 
(ii)~\textit{Cold-start Optimization Strategies}, describing optimization and reengineering practices; and 
(iii)~\textit{Localization Challenges}, summarizing diagnostic barriers reported by developers. 
Coding continued until \emph{theoretical saturation}. 
Inter-rater reliability was substantial (Cohen’s~$\kappa{=}0.89$), consistent with prior taxonomy studies~\cite{rahman2020gang,kim2021studying,humbatova2020taxonomy}, and discrepancies were later resolved through discussion. 
The final taxonomy includes six cold-start inefficiency themes, six remediation strategies, and five recurrent localization-challenge categories.

\subsection{Phase II: \bench{} Microbenchmark Construction}
\label{sec:phase2}
We develop \emph{Serverless Cold-Start Analysis Benchmark (\bench{})}, following established benchmark guidelines~\cite{lu2005bugbench,kashyap2019buginjector,gray1991benchmark} to provide small yet representative functions that isolate initialization inefficiencies under reproducible, tool-independent conditions.

\textbf{Benchmark contents.}
The current release of \bench{} includes 18 scenarios derived from the five localization challenge categories.  
Each scenario reflects one cross-layer inefficiency observed in the mined issues, such as redundant dependency loading (design), nested packaging overheads (packaging), handler-scope initialization (runtime), or environment-dependent credential lookup (environment).  
Every case provides a minimal source package, a driver that triggers the inefficiency, and a metadata file describing its ground-truth layer, affected components, and initialization phase, following composition practices of prior benchmarks~\cite{lu2005bugbench,kashyap2019buginjector}.

\textbf{Design rationale.}  
\bench{} adopts a synthetic yet empirically informed design that captures representative initialization behavior rather than replicating full applications.  
Each scenario defines explicit ground truth for the impacted layer and component, enabling later evaluation of localization and attribution accuracy across tools.

\textbf{Scenario structure.}  
Each variant consists of a small source tree, an entry module, and a callable function that executes the target path. A test harness validates scenario preconditions and deterministically triggers the target path before measurement, ensuring reliable ground truth, as recommended by prior benchmarks \cite{lu2005bugbench,just2014defects4j,madeiral2019bears}.  
Scenario metadata specifies the ground-truth layer, expected phase, and attribution tags (\texttt{must\_blame}/\texttt{must\_not\_blame}) for consistent comparison.

\textbf{Parameterization and realism.}  
Scenarios expose tunable parameters (e.g., import fan-out, payload size, number of files) that scale initialization cost without altering semantics.  
Where native boundaries are modeled, both simulated (CPU sleep or loop) and real variants (e.g., \texttt{ctypes} into \texttt{libsqlite3}) are provided, following representativeness and variability principles~\cite{gray1991benchmark}.

\textbf{Reproducibility and Benchmark Stability.}  
All benchmarks execute in isolated Docker containers to ensure deterministic environments and eliminate host interference.  
Each scenario was executed multiple times to verify stable latency patterns and consistent cost ratios, with overall variance remaining below 5\%.  
A 20\% subset of scenarios was re-run on alternate hosts, yielding median differences within $\pm$3\%, confirming platform independence.  
Benchmark construction is independent of the analysis tools evaluated later, minimizing bias~\cite{kashyap2019buginjector}.

\subsection{Phase III: Profiling and Comparative Evaluation on \bench{}}
\label{sec:phase3}

This phase quantitatively evaluates cold-start inefficiencies using our profiling instrument, \tool{}, and compares its localization accuracy against existing analysis tools on the controlled scenarios from \bench{}.

\textbf{Instrumentation and design.}
\tool{} implements a hybrid profiling architecture that captures both import-time and runtime behavior to attribute initialization latency precisely. 
It instruments Python’s \texttt{importlib} subsystem to intercept module load events, record timestamps, and extract import stacks, producing a fine-grained \emph{import-phase trace}. 
A lightweight sampling agent concurrently collects call stacks during warm executions to estimate invocation frequency for each imported component. 
Both traces are merged by aligning module identifiers and call contexts to construct an annotated Calling-Context Tree (CCT)~\cite{ammons1997exploiting} encoding cumulative initialization latency, post-load usage frequency, and inter-module dependencies. 
This correlation of import-time latency with execution frequency bridges static dependency analysis and dynamic profiling, revealing modules that are costly to load yet rarely used.

\textbf{Analysis and visualization.}
Beyond profiling, \tool{} integrates an interactive visual analyzer that distinguishes it from black-box profilers such as SlimStart. 
The analyzer renders merged CCTs as multi-view dashboards: (i)~an \emph{Initialization Overhead View} ranking modules by cold-start cost; 
(ii)~an \emph{Inefficiency Prioritization View} combining initialization cost with usage frequency; and 
(iii)~a \emph{Source-Level Context View} mapping delays and usage directly to source lines and import statements. 
Developers can filter modules by layer or sort by inefficiency score to focus on the most critical imports, translating profiling data into actionable refactoring guidance.

\textbf{Execution protocol.}
Each \bench{} scenario and 22 real-world serverless applications were executed in a controlled environment (Python~3.9, 512~MB memory). 
Configurations were run for 500 cold starts over five repetitions with randomized order and container resets to remove caching or network bias. 
Results were aggregated using median-of-medians statistics following established profiling practices~\cite{ammons1997exploiting,slimstart}. 
Profiling overhead remained below 10\% when compared with non-instrumented baseline runs across both SCABENCH benchmarks and real applications, validating the lightweight design of \tool{}. 
All calibration data and measurement scripts are included in the replication package for reproducibility.

\textbf{Evaluation baselines.}
We compared three workflows: 
(a)~\emph{FaaSLight}, a static-analysis slimming baseline; 
(b)~\emph{SlimStart}, a profile-guided slimming baseline; and 
(c)~\emph{\tool{}}, our profiling-based attribution workflow. 
Each benchmark encodes explicit ground truth for the root-cause module and layer, enabling consistent evaluation across tools.

\textbf{Metrics and analysis.}
We report end-to-end cold-start latency as well as per-import initialization time. 
For every import site, \tool{} records initialization time, usage count, and call-depth context. 
A usage-normalized inefficiency score,
$U=\tfrac{\text{InitTime}}{\max(1,\text{UsageCount})}$,
highlights costly but seldom-used components and serves as a proxy for optimization opportunity. 
Localization quality was assessed using \texttt{precision}, \texttt{recall}, and false-positive rate (removal of required imports). 
Statistical significance was evaluated with Wilcoxon signed-rank and Mann–Whitney~U tests with Holm–Bonferroni correction; effect sizes were reported via Cliff’s~$\delta$ with 95\% confidence intervals.

\subsection{Phase IV:  Developer User Study}
\label{sec:phase4}

This phase examines how diagnostic feedback influences developers’ ability to identify and reason about initialization inefficiencies. The study protocol received institutional ethics approval from our university’s review board (IRB). All participants were informed of the study purpose, procedures, and data-handling policy, and provided written consent in accordance with the approved protocol.

\textbf{Participants and recruitment.}
We recruited six professional software engineers (three industry practitioners, three experienced open-source contributors) familiar with Python and serverless deployment.
All participation was voluntary, with informed consent obtained ,and no monetary compensation.

\textbf{Design and tasks.}
We employed a within-subject, repeated-measures design with counterbalanced order.
Each participant completed four tasks on a real serverless application:  
T1: detect costly imports; T2: localize the import site;  
T3: classify the inefficiency using Phase II criteria; and  
T4: propose an optimization.  
Each participant performed the sequence first without, then with \tool{}, with a 20-minute cap per task.

\textbf{Measures and procedure.}
We recorded task completion time and correctness, followed by short reflections.
Participants received a ten-minute orientation to the diagnostic interface.
Sessions were screen-recorded (with consent) for timing verification and qualitative notes.
No surveys were administered; analysis is based on observed behavior and task outcomes.

\textbf{Analysis.}
Performance differences between conditions were evaluated using non-parametric tests with effect sizes and 95\% confidence intervals.
Qualitative observations were coded to identify recurring reasoning strategies and localization challenges, then triangulated with results from Phases I-III.

%% file: sections/4-Evaluation.tex
\section{Evaluation}
\label{eval}
This section evaluates our results along the four research questions introduced in Section~\ref{methodology}: 
\textbf{(RQ1)} which \emph{initialization anti-patterns} arise in practice and how they distribute across layers; 
\textbf{(RQ2)} which \emph{refactoring strategies} remediate each anti-pattern and how they compose; 
\textbf{(RQ3)} which \emph{diagnostic barriers} impede localization; and \textbf{(RQ4)}

\subsection{\textbf{RQ1: Which initialization \emph{anti-patterns} occur in practice?}}
\label{sec:rq1}

\begin{figure}[t]
  \centering
  \includegraphics[width=\columnwidth]{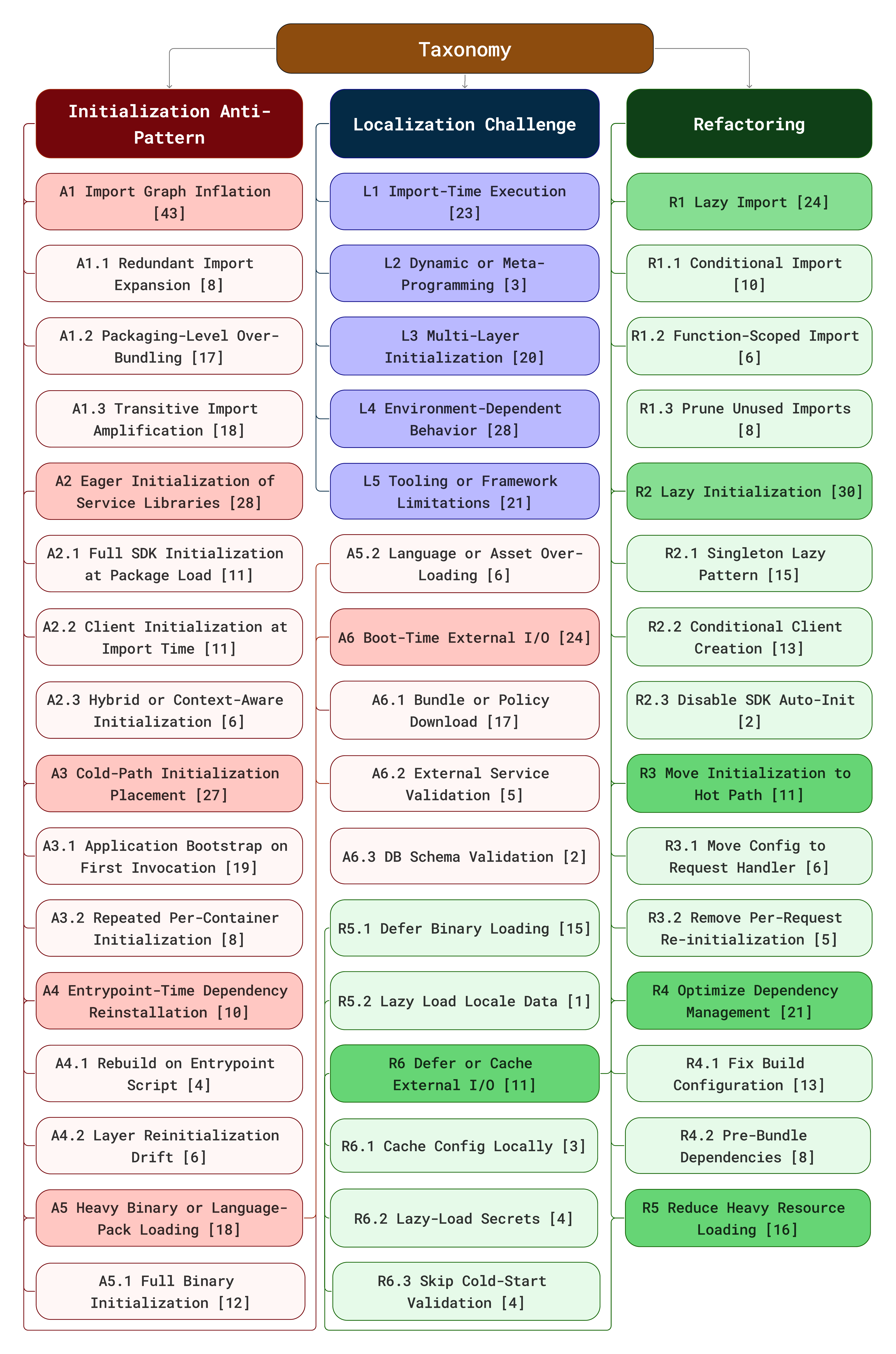}
  \caption{Unified taxonomy: initialization anti-patterns (A1–A6), refactoring strategies (R1–R6), and localization challenges (L1–L5).}
  \label{fig:unified_taxonomy}
\end{figure}

We derive a \emph{root-cause taxonomy} of initialization anti-patterns from 81 adjudicated issue reports in open-source serverless and cloud-native projects. 
Each anti-pattern is defined by its \emph{root cause}: the mechanism introducing inefficiency, and its \emph{system layer of occurrence}, spanning (\emph{i})~\textbf{Design}, (\emph{ii})~\textbf{Packaging}, (\emph{iii})~\textbf{Runtime}, and (\emph{iv})~\textbf{Environment}. 
We focus on developer-reported mechanisms, using qualitative evidence rather than quantitative metrics. Figure~\ref{fig:unified_taxonomy} summarizes the taxonomy.

We applied a strict labeling policy, assigning subcategories only when explicit evidence appeared in issue discussions. 
Out of 81 analyzed issues, 72 met this criterion. 
Each anti-pattern family (A1–A6) comprises multiple recurring sub-themes, such as redundant imports, over-bundled dependencies, and boot-time network operations, described below.

\medskip
\noindent\textbf{A1: Import Graph Inflation (Design, Packaging, Runtime)} \textit{(n=\textbf{43})}  
Top-level imports, bundler defaults, or nested artifacts expand dependency graphs into the cold path. 
This includes redundant imports that are never used, over-bundled packages that serialize entire applications, and transitive import chains that unnecessarily pull secondary modules. 
Developers frequently link these patterns to inflated memory usage and cold-start latency.

\noindent\textbf{A2: Eager Initialization of Service Libraries (Runtime)} \textit{(n=\textbf{28})}  
Service libraries often initialize eagerly, either through SDKs performing static setup at package load or developer code instantiating clients at import time (e.g., \texttt{boto3.client('s3')}). 
In some cases, initialization mixes import-time and runtime configuration, compounding cold-start delay through premature credential or network operations.

\noindent\textbf{A3: Cold-Path Initialization Placement (Runtime, Container)} \textit{(n=\textbf{27})}  
One-time setup logic such as ORM registration, schema compilation, or application bootstrap often executes on the request-critical path or repeats per container. 
This misplaced initialization defers heavy setup work to first invocation, prolonging cold starts and causing redundant container-level reinitialization.

\noindent\textbf{A4: Entrypoint-Time Dependency Reinstallation (Environment, Cross-Layer)} \textit{(n=\textbf{10})}  
Containers frequently reinstall or rebuild dependencies at startup due to misconfigured entrypoint scripts or layer drift, where cached dependencies are reconstructed because of version mismatches or stale metadata. 
These behaviors lead to recurring startup overheads even for otherwise static deployments.

\noindent\textbf{A5: Heavy Binary or Language-Pack Loading (Packaging, Runtime)} \textit{(n=\textbf{18})}  
Large native binaries and full language packs are often loaded eagerly, creating high I/O and CPU demand. 
Examples include headless browser libraries or preloaded assets (e.g., grammar packs, themes) that could instead be loaded lazily. 
Such over-loading amplifies startup cost without proportionate runtime benefit.

\noindent\textbf{A6: Boot-Time External I/O (Runtime, Environment)} \textit{(n=\textbf{24})}  
Many applications perform blocking external I/O during startup, such as downloading policy bundles, validating external services, or fetching credentials, before any request is processed. 
These boot-time dependencies, especially when network-bound, introduce the longest and most unpredictable cold-start delays.

\noindent\textbf{Unlabeled Cases.}  
Nine issues (11\%) could not be confidently classified under our strict criteria and describe general startup delays or infrastructure effects without clear application-level mechanisms.

\subsubsection*{Distribution and co-occurrence}

Across the 72 labeled issues, three dominant clusters emerge. 
(\emph{i}) \textbf{Scope expansion} (A1, A5) remains the most common, where import graphs, bundles, or binary footprints grow excessively. 
(\emph{ii}) \textbf{Initialization placement} (A2, A3) follows closely, capturing cases where one-time setup work executes in an inappropriate phase or context. 
Together, these four families account for nearly 90\% of all classified issues, showing that most initialization inefficiencies stem from redundant dependency loading or misplaced initialization logic. 
\textbf{A1} (\textbf{43} cases) dominates the corpus, followed by \textbf{A2} (\textbf{28}) and \textbf{A3} (\textbf{27}), while \textbf{A6} (\textbf{24}) also appears frequently, often in combination with A1 and A2. 
Environment-driven mechanisms (\textbf{A4}, \textbf{A6}) are less common overall (about \textbf{30\%} combined) but contribute the most persistent startup latencies, such as container rebuilds and boot-time network calls.

\begin{figure}[t]
  \centering
  \includegraphics[width=0.75\linewidth]{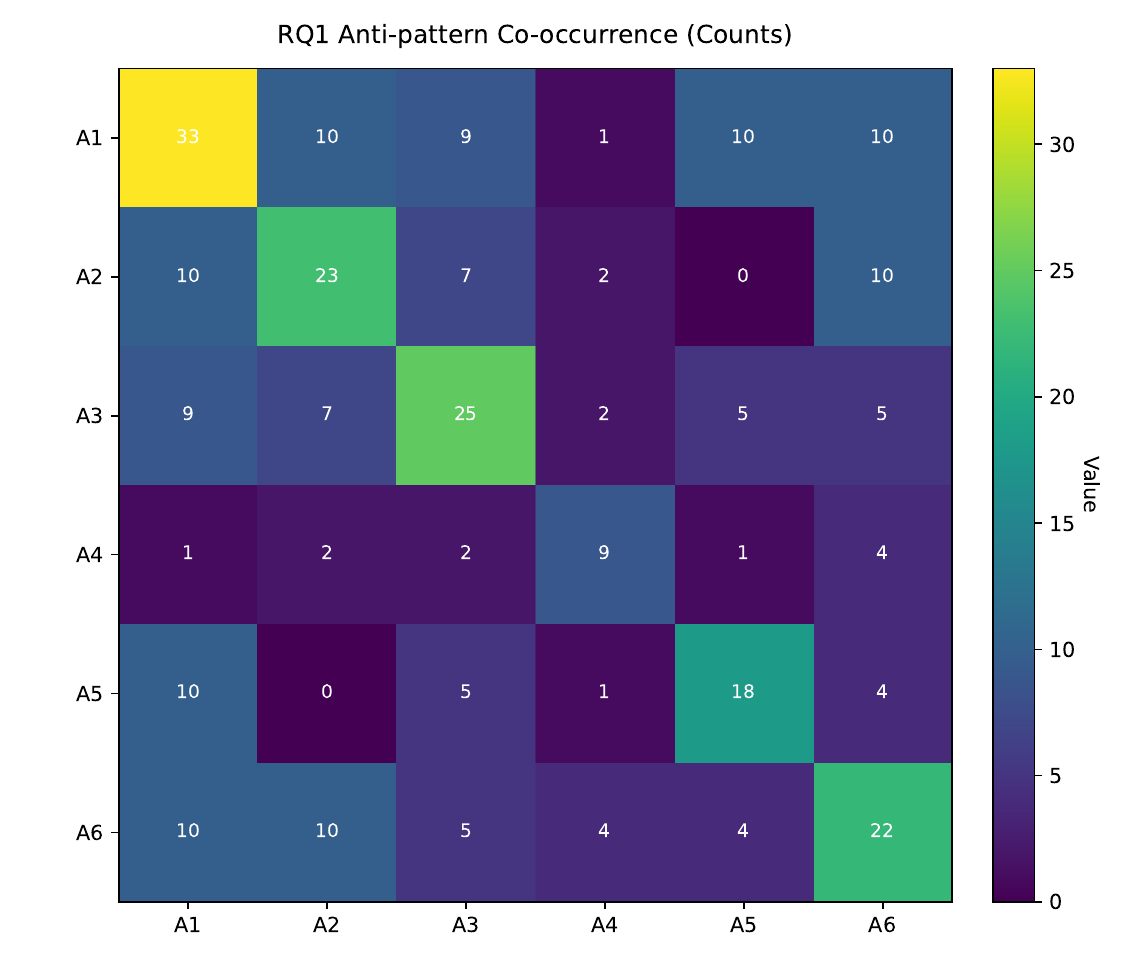}
  \caption{Co-occurrence heatmap of initialization anti-patterns.}
  \label{fig:antipattern_heatmap}
\end{figure}

Co-occurrence analysis in Figure~\ref{fig:antipattern_heatmap} highlights a prominent triad among \textbf{A1 (Import Graph Inflation)}, \textbf{A2 (Eager Service Initialization)}, and \textbf{A6 (Boot-Time External I/O)}, with each pair appearing together in ten issues. This combination suggests that dependency scope, initialization timing, and external I/O frequently interact, forming a recurring cross-layer pattern of cold-start overhead. Examples include \textit{AWS Powertools} and \textit{Zappa}, where top-level imports (e.g., \texttt{boto3.client('s3')}) initialize SDK clients during module load, and \textit{Open Policy Agent} and \textit{Nitro}, where large bundles combine with external configuration downloads at boot. Moderate links between \textbf{A1}, \textbf{A3 (Cold-Path Initialization Placement)}, and \textbf{A5 (Heavy Binary or Language Pack Loading)} (9-10 overlaps) suggest that dependency inflation often coincides with misplaced initialization and packaging overheads. Overall, initialization inefficiencies tend to cascade across dependency, initialization, and I/O layers, underscoring the need for holistic analysis across design, packaging, and runtime dimensions.

\begin{customquote}
\textbf{Finding 1.}
\textit{Over 80\% of observed inefficiencies stem from two primary sources: 
excessive dependency loading and misplaced initialization. 
The most frequent interactions occur when broad import graphs coincide with premature service initialization or startup-time network activity, 
revealing a strong coupling between dependency scope and runtime behavior. 
Environment-driven cases are less common overall but are often described as the most persistent or complex startup problems.}
\end{customquote}

\subsection{\textbf{RQ2: Which \emph{refactoring strategies} remediate these anti-patterns, and how are they applied?}}
\label{sec:rq2}

From the 81 adjudicated issues, we identified 64 cases (79\%) that proposed explicit optimization strategies. 
These form an evidence-based taxonomy of refactorings observed in practice.

\noindent\textbf{R1: Lazy Import (Runtime)} \textit{(n=\textbf{24})}  
Developers defer module loading until first use or guard imports behind runtime conditions. 
Typical changes include moving imports into function scope, pruning unused modules, or conditionally loading SDK helpers and validators to reduce import-time overhead.

\noindent\textbf{R2: Lazy Initialization (Runtime)} \textit{(n=\textbf{30})}  
Initialization of clients, services, or SDKs is postponed until actually needed. 
This includes introducing singleton patterns, conditional client creation, or disabling automatic initialization logic within libraries to shorten the cold-start path.

\noindent\textbf{R3: Move Initialization to Hot Path (Application Logic)} \textit{(n=\textbf{11})}  
One-time setup work is relocated from startup to the request handler or cached between invocations. 
This approach prevents repeated reconfiguration or reinitialization on every cold start.

\noindent\textbf{R4: Optimize Dependency Management (Packaging/Build)} \textit{(n=\textbf{21})}  
Build configurations and dependency graphs are refined to reduce load time and deployment size. 
Common fixes include correcting build metadata, pre-bundling dependencies, or restructuring modules to limit unnecessary packaging overhead.

\noindent\textbf{R5: Reduce Heavy Resource Loading (Runtime/Resources)} \textit{(n=\textbf{16})}  
Large binaries or resource packs are loaded only when required. 
This includes deferring binary initialization, lazy-loading locale data, and removing non-essential assets from startup.

\noindent\textbf{R6: Defer or Cache External I/O (Environment)} \textit{(n=\textbf{11})}  
External lookups and validation steps are deferred, cached, or skipped when safe. 
Typical practices include caching configuration locally, lazy-loading secrets, or bypassing repeated network validations to minimize environment-induced latency.

\subsubsection*{Distribution of remediation strategies}

Table~\ref{tab:anti_refactor_mapping} summarizes how inefficiency patterns are mitigated in practice. 
Most fixes aim to shrink the cold-path footprint: developers either defer imports and client setup until first use or refine build/dependency configuration so that less is packaged and parsed at startup. 
A smaller but still substantial set of changes addresses initialization behavior itself by relocating one-time setup into the request path or eliminating repeated work. 
Finally, developers reduce resource pressure and environment costs by deferring large binaries and locale data, caching configuration locally, or skipping redundant external checks when safe.

Import-graph inflation (A1) most often co-occurs with deferring imports and tightening dependency/build configuration; eager or misplaced initialization (A2-A3) aligns with postponing client setup and moving initialization out of startup; heavy binaries (A5) aligns with reducing resource loading; and environment-coupled startup I/O (A6) aligns with postponing initialization and optimizing dependency/build configuration. 
Because issues frequently apply multiple remedies, totals in Table~\ref{tab:anti_refactor_mapping} reflect co-occurrence intensity rather than unique counts.

\begin{customquote}
\textbf{Finding 2.}
\textit{Remediations cluster into three complementary approaches. 
Developers commonly minimize dependency and packaging scope through deferred imports, bundle pruning, and modular packaging. 
They also optimize initialization behavior by introducing caching mechanisms and precompiling schemas or configuration data. 
Finally, they stabilize runtime environments through prebuilt artifacts, lightweight containers, and targeted configuration tuning. 
Together, these approaches span design, packaging, and environment layers, illustrating how mitigation practices extend from code-level reduction to cross-layer optimization.}
\end{customquote}


\begin{table}
\centering
\caption{Mapping between initialization anti-patterns (rows) and refactoring families (columns) under strict co-label evidence.}
\label{tab:anti_refactor_mapping}
\begin{tabular}{lrrrrrrr}
\toprule
{} & R1 & R2 & R3 & R4 & R5 & R6 & Total \\
\midrule
A1 & 16 &  9 &  1 & 11 &  8 &  1 & 46 \\
A2 &  4 & 12 &  5 &  3 &  1 &  3 & 28 \\
A3 &  7 & 10 &  8 &  5 &  3 &  3 & 36 \\
A4 &  0 &  1 &  1 &  6 &  4 &  2 & 14 \\
A5 &  5 &  5 &  0 &  5 & 11 &  2 & 28 \\
A6 &  3 &  8 &  4 &  6 &  4 &  4 & 29 \\
\bottomrule
\end{tabular}
\end{table}

\subsection{\textbf{RQ3: Which \emph{localization challenges} impede attribution, and how do they manifest?}}
\label{sec:rq3}

We identified localization challenges in 64 of 81 issues (79\%) that explicitly described diagnostic or attribution difficulties. 
Labels were assigned only when supported by concrete evidence, and multiple categories could apply to a single issue.

\noindent\textbf{L1: Import-Time Execution} \textit{(n=\textbf{23})}  
Initialization often occurs implicitly at import time, making performance bottlenecks invisible in typical control flow.  
Developers struggle to attribute delays because costly operations are triggered by hidden module-level code or nested re-exports rather than explicit runtime logic.

\noindent\textbf{L2: Dynamic or Meta-Programming} \textit{(n=\textbf{3})}  
Reflection, decorators, and dynamic code generation introduce layers of indirection that mask where initialization work occurs.  
Such mechanisms complicate profiling because initialization is intertwined with runtime adaptation or dependency injection.

\noindent\textbf{L3: Multi-Layer Initialization} \textit{(n=\textbf{20})}  
Initialization behavior frequently spans middleware, ORM, secrets management, and networking layers, dispersing cold-path costs across components.  
This entanglement makes it difficult to determine which layer is responsible for startup latency or redundant setup work.

\noindent\textbf{L4: Environment-Dependent Behavior} \textit{(n=\textbf{28})}  
Cold-start latency varies with runtime configurations, container settings, or network conditions, leading to inconsistent and irreproducible measurements.  
Diagnosing these issues often requires environment-specific experiments, as performance anomalies disappear outside the affected context.

\noindent\textbf{L5: Tooling or Framework Limitations} \textit{(n=\textbf{21})}  
Available profilers and tracing tools provide incomplete visibility into initialization behavior within complex frameworks or compiled modules.  
Developers frequently face blind spots where crucial initialization steps are hidden behind framework abstractions or missing instrumentation.

\begin{table}[t]
\centering
\caption{Mapping between initialization anti-patterns (rows) and localization challenges (columns) under strict co-label evidence.}
\label{tab:a_l_mapping_relabel}
\begin{tabular}{lrrrrrr}
\toprule
{} & L1 & L2 & L3 & L4 & L5 & Total \\ 
\midrule
A1 & 16 & 1 & 7 & 13 & 12 & 49 \\
A2 & 12 & 0 & 10 & 11 & 4 & 37 \\
A3 & 10 & 0 & 13 & 8 & 6 & 37 \\
A4 & 2 & 0 & 2 & 3 & 4 & 11 \\
A5 & 5 & 1 & 3 & 6 & 8 & 23 \\
A6 & 5 & 1 & 4 & 9 & 6 & 25 \\
\bottomrule
\end{tabular}
\end{table}

\subsubsection*{Distribution and implications}

Four categories dominate the dataset: environment-dependent behavior (\textbf{L4}), import-time execution (\textbf{L1}), multi-layer initialization (\textbf{L3}), and tooling or framework limitations (\textbf{L5}). 
Together they account for nearly 90\% of all reported localization challenges, confirming that cold-start inefficiencies are inherently cross-layer rather than confined to code-level logic. 
As summarized in Table~\ref{tab:a_l_mapping_relabel}, the strongest co-occurrences occur between \textbf{A1/A2} and \textbf{L1/L4}, indicating that dependency inflation and eager initialization frequently manifest as hidden import work and environment-sensitive startup delays. 
Multi-layer initialization (\textbf{L3}) often aligns with misplaced initialization patterns (\textbf{A2-A3}), while tooling limitations (\textbf{L5}) are distributed across nearly all anti-pattern families. 
These overlaps illustrate that the same structural complexity that causes cold-start inefficiencies also obscures their detection, underscoring the need for holistic, environment-aware diagnostic approaches.

\begin{customquote}
\textbf{Finding 3.}
\textit{Localization barriers are dominated by environment-dependent behavior, import-time execution, multi-layer initialization, and tooling limitations. 
These challenges highlight that cold-start inefficiencies emerge from interactions across code, framework, and environment layers, demanding cross-layer, context-aware diagnostic approaches.}
\end{customquote}

\subsection{\textbf{RQ4: Effectiveness of Analysis Tools}}

This section evaluates how effectively existing tools detect and localize cold-start inefficiencies. We first perform a quantitative assessment on \bench{}, comparing \tool{} with two state-of-the-art baselines, followed by a qualitative user study examining its practical utility and developer acceptance in real diagnostic workflows.

\subsubsection{\textbf{Quantitative Evaluation of State-of-the-art Tools}}
\label{quantiitative_evaluation}
\input{sections/5-quantitative_analysis}

\subsubsection{\textbf{Qualitative Evaluation of \tool{}}}
\label{qualitative_evaluation}

\input{sections/6-user-study-results}

%% file: sections/5-quantitative_analysis.tex
To address \textbf{RQ4}, we quantitatively evaluate how accurately \tool{} detects and localizes cold-start inefficiencies against two state-of-the-art baselines: FaaSLight~\cite{liu2023faaslight}, a static reachability analyzer, and SlimStart~\cite{slimstart}, a profile-guided optimizer. Experiments followed a uniform protocol on \bench{}~\ref{sec:phase2}, executed on a CloudLab \texttt{xl170} bare-metal node (Ubuntu~24.04) in isolated Docker containers for reproducibility. 

\bench{} comprises 18 Python micro-benchmarks spanning eight cold-start patterns (B1–B8), each with explicit ground-truth attribution for the root-cause modules. Categories include import-graph indirection, transitive dependency dominance, import-time side effects, deferred initialization, loader overheads, cross-language initialization, framework discovery, and resource-loading policy. Each tool was evaluated across 20 repetitions, measuring Precision, Recall, F1, and false-positive/negative counts.

FaaSLight’s static analysis attains high recall but low precision due to over-attribution of reachable code, while SlimStart improves precision via runtime sampling but misses loader and native-boundary cases. \tool{} correlates import-time latency with subsequent runtime usage, assigning higher weight (0.8) to latency and lower weight (0.2) to usage. This hybrid scoring highlights modules that are costly to load yet seldom used, improving detection precision. 

\begin{table}[t]
\centering
\scriptsize
\setlength{\tabcolsep}{3pt}
\renewcommand{\arraystretch}{1.15}
\begin{tabular}{@{}p{0.5\columnwidth} c c c@{}}
\toprule
\textbf{Benchmark} & \textbf{FaaSLight} & \textbf{SlimStart} & \textbf{\tool{}} \\
\midrule
B1 - Import-Graph Indirection & \fullmark & \fullmark & \fullmark \\
B2 - Transitive Dependency Dominance & \partialmark & \fullmark & \fullmark \\
B3 - Import-Time Side Effects & \partialmark & \fullmark & \fullmark \\
B4 - Deferred First-Use Initialization & \fullmark & \fullmark & \fullmark \\
B5 - Loader and Packaging Overheads & \fullmark & \partialmark & \fullmark \\
B6 - Cross-Language Boundary & \partialmark & \nonemark & \nonemark \\
B7 - Framework Discovery Scan & \fullmark & \fullmark & \fullmark \\
B8 - Resource Loading Policy & \fullmark & \fullmark & \fullmark \\
\midrule
\multicolumn{4}{@{}l}{\textbf{Tool Performance Metrics}} \\
\multicolumn{1}{@{}l}{True positive (TP), higher is better} & \textbf{15} & \textbf{13} & \textbf{16} \\
\multicolumn{1}{@{}l}{False positive (FP), lower is better}            & \textbf{21} & \textbf{4}  & \textbf{3}  \\
\multicolumn{1}{@{}l}{Precision $p=$\textit{\scriptsize TP/(TP+FP)}}                 & \textbf{41.7\%} & \textbf{75.0\%} & \textbf{78.0\%} \\
\multicolumn{1}{@{}l}{Recall $r=$\textit{\scriptsize TP/(TP+FN)}}                    & \textbf{88.2\%} & \textbf{70.6\%} & \textbf{82.4\%} \\
\multicolumn{1}{@{}l}{F1-score $=2pr/(p+r)$}                    & \textbf{56.6\%} & \textbf{72.7\%} & \textbf{80.0\%} \\
\bottomrule
\end{tabular}
\caption{Comparison of existing tools using \bench{}. $\checkmark$ denotes full success, $\times$ denotes full miss, and \partialmark\ denotes partial localization success (mixed outcomes).}
\label{tab:scabench-compact}
\end{table}

\textbf{Table~\ref{tab:scabench-compact}} summarizes results across all benchmark categories. \tool{} consistently outperforms both baselines, achieving \textbf{78.0\% precision}, \textbf{82.4\% recall}, and \textbf{80.0\% F1}, while reducing false positives by over 85\% relative to FaaSLight. These gains, statistically significant (\emph{p}~$<$~0.01, Cliff’s~$\delta$~$>$~0.47), confirm the benefit of hybrid import–execution correlation for explainable cold-start localization.

\begin{customquote}
\textbf{Finding 4.}
\textit{While static analyzers such as FaaSLight achieve high recall and dynamic profilers such as SlimStart reduce noise, both function as black boxes with limited diagnostic value. \tool{} bridges this gap through hybrid import-execution correlation and visual attribution, providing high accuracy and actionable insight that help developers localize and optimize cold-start inefficiencies.}
\end{customquote}

%% file: sections/6-user-study-results.tex
This section reports the findings of our qualitative \emph{user study} (see~\ref{sec:phase4}), which evaluated \tool{}’s effectiveness in identifying and localizing cold-start inefficiencies in real serverless applications. Six participants, software engineers and researchers with 1-20 years of experience (1-10 in serverless development), completed four hands-on tasks under two conditions (manual baseline and with \tool{}). We measured completion time and accuracy and analyzed screen recordings and notes to capture reasoning strategies and workflow impact.

\begin{customquote}
\textbf{Finding 5 (Developer Study):} \tool{} substantially improved developer productivity, reducing detection and localization time by 64\% and 43\%, respectively, while increasing task accuracy from 14\% to 71\%.
\end{customquote}

In the manual baseline condition, over 80\% of participants failed to accurately detect or localize cold-start inefficiencies and spent considerably more time per task. With \tool{}, completion times decreased across detection (T1), localization (T2), classification (T3), and optimization (T4) by 59.7\%, 37.7\%, 64.2\%, and 43.9\%, respectively. Qualitative observations indicated that \tool{} helped participants recognize import-time costs and dependency hot spots earlier, enabling faster and more confident diagnostic reasoning.

Participants praised \tool{}’s visualization and scoring features for reducing manual search effort and improving localization accuracy. One noted, \emph{“\tool{} directly gives the code file using the library, reducing unnecessary search time”} (P5). Others suggested adding automated optimization guidance and improved scalability, for example, \emph{“\tool{} should suggest optimizations automatically and may integrate Gen AI to enable this feature”} (P2, General Manager). These insights point to opportunities for AI-assisted recommendations and integration with continuous-development workflows.

%% file: sections/7-threats-to-validity.tex
\section{Discussion and Implications}
\label{discussion}

Our findings frame initialization inefficiencies as a \textit{system-level coordination problem} linking software design, build automation, and runtime behavior. These inefficiencies reveal a deeper misalignment between how developers intend initialization to operate and how systems actually execute it, carrying implications for researchers, practitioners, and tool designers in performance-aware software engineering.

\subsection{Implications for Researchers}
The three taxonomies collectively uncover recurring cross-layer mechanisms and reasoning strategies that shape cold-start inefficiencies. These mechanisms resemble \textbf{cross-layer technical debt} that accumulates when initialization responsibilities drift across abstraction boundaries. Such debt is architectural rather than syntactic, manifesting as temporal coupling between build, runtime, and environment layers. Future research can model how this debt evolves, co-occurs with architectural decay, and propagates through refactoring cycles. Our results also emphasize the need to integrate architectural recovery, socio-technical congruence~\cite{Cataldo2008SocioTech}, and program-comprehension theory~\cite{pirolli1999information,lawrance2008information} to better explain how developers reason about performance beyond numerical metrics.

\subsection{Implications for Practitioners and Architects}
From an engineering perspective, initialization should be managed as an explicit architectural concern rather than an incidental runtime effect. Developers often design for modularity but not for \textit{initialization boundaries}, producing hidden coupling that expands dependency scope and startup cost. Our analysis supports introducing \textbf{initialization contracts}: declarative specifications of where, when, and under what context dependencies are loaded. Such contracts, embedded in configuration metadata or build manifests, can make initialization behavior traceable and enforceable. Architects can extend principles of fault containment and interface design to isolate initialization responsibilities across system layers, improving both observability and reuse.

\subsection{Implications for Tool Builders and Framework Designers}
Existing profilers and optimizers often treat initialization as a black-box phenomenon, reporting latency without explaining causality. Our results highlight a missing \textbf{semantic continuity} between static dependency graphs and dynamic execution traces. Tools such as \tool{} demonstrate that linking \textit{what is loaded} with \textit{what is executed} provides actionable insight into cold-path behavior. Future diagnostic frameworks should merge import-time tracing with execution sampling to support explainable attribution of initialization cost. Similarly, framework designers should expose explicit hooks and APIs for managing cold-start phases, allowing developers to control and monitor initialization declaratively.

\subsection{Toward a Theory of Initialization Debt}
Across the three perspectives, anti-patterns, refactorings, and localization challenges, our results suggest a distinct form of \textbf{initialization debt}, or \textit{cold-start debt} in serverless contexts. 
Unlike conventional technical debt from postponed refactoring, initialization debt arises when startup responsibilities drift across design, packaging, and environment layers without clear ownership or visibility. 
This debt is both cognitive and architectural: developers often assume performance costs are confined to runtime logic, leaving import- and build-time costs unexamined, a mismatch between mental models and execution reality. 
Addressing this debt requires shifting performance reasoning upstream and treating initialization behavior as a first-class architectural boundary.

\subsection{Broader Reflection and Future Outlook}
Initialization inefficiencies exemplify a broader class of \textbf{latent performance dependencies} that cross traditional abstraction boundaries. Our mixed-method design, combining repository mining, controlled benchmarking, and developer studies, shows how performance can be studied as a traceable and explainable software property. Extending this approach to other performance domains, such as dependency caching or container reuse, can advance an empirical foundation for treating performance as an observable and governable aspect of software architecture.

\section{Threats to Validity}
\label{threats}

\noindent\textbf{Internal Validity.}  
Our findings depend on the accuracy of the methods used for issue mining, benchmarking, and the developer study.  
We mitigated potential confounds by manually validating mined issues, executing benchmarks under identical conditions, and training participants before experimentation.  
The within-subject, counterbalanced design helped reduce learning effects, although residual carry-over between conditions may still exist. The LLM-assisted triage during issue mining could introduce classification bias; this was mitigated through manual verification by two experienced raters with substantial agreement (Cohen's~$\kappa{=} 0.84$). Longer-term studies in industrial settings would further strengthen these results.

\noindent\textbf{Construct Validity.}  
Threats concern whether our instruments correctly capture initialization inefficiencies and diagnostic behavior.  
The developer tasks were derived directly from the taxonomies identified in Phase~I, aligning observed behavior with the studied constructs.  
Still, the controlled environment may simplify real-world workflows.  
Future replications could employ field studies to observe natural diagnostic processes and contextual influences.

\noindent\textbf{External Validity.}  
Our evaluation focused on Python-based serverless systems, where module-import semantics dominate cold-start cost.  
Although the conceptual framework is language-agnostic, other ecosystems (e.g., Java, Node.js) may exhibit different runtime or build characteristics.  
Extending \tool{} and \bench{} to multiple languages and platforms will further confirm the generality of our findings.

\noindent\textbf{Conclusion Validity.}  
We mitigated random variation through repeated measurements, containerized execution, and non-parametric statistical tests with Holm-Bonferroni correction.  
While profiling overhead and cloud resource variability may affect absolute latency, our analysis focuses on relative comparisons to ensure robust inference.  
Replication with larger participant pools and diverse environments would enhance statistical power and confirm our conclusions.

%% file: sections/8-Conclusions.tex
\section{Conclusions}
\label{conclusions}
This study shows that cold-start inefficiencies are systemic, cross-layer phenomena arising from the interplay of design, packaging, and runtime behavior in serverless systems. By deriving empirical taxonomies of anti-patterns and refactorings, instantiating them in \bench{}, and validating diagnostic improvements with \tool{}, we offer both an explanatory foundation and practical instrumentation for cold-start analysis. Unifying cause, remediation, and diagnosis perspectives transforms cold-start optimization from ad hoc tuning into an evidence-driven engineering practice. Future work will extend this cross-layer lens to other ecosystems and programming models, promoting reproducible, developer-centered approaches to performance reasoning in cloud-native software.

\section{Data Availability}
\label{data-availability}
The research artifact is publicly accessible for future studies and improvements at \href{https://doi.org/10.5281/zenodo.17375894}{\texttt{https://doi.org/10.5281/zenodo.17375894}}.

%% file: sections/9-references.tex
\bibliographystyle{IEEEtran}
\bibliography{BibFile}